\documentclass[11pt]{article}
\usepackage{geometry}
\usepackage{here}
\usepackage{graphicx}
\usepackage{amsmath,amssymb}
\usepackage{cite}
\usepackage{bm}
\newcommand{\lw}[1]{\smash{\lower 1.5ex\hbox{#1}}}
\newcommand{\dint}[0]{\displaystyle{\int_0^{\infty}}}
\newcommand{\mapright}[1]{\smash{\mathop{\hbox to 1cm{\rightarrowfill}}\limits^{#1}}}

\begin{document}

\title{
Analysis of the L3 BEC at Z$^0$-pole\\
{\Large -- Comparison of the conventional formula against the $\tau$-model --}}

\author{Takuya Mizoguchi$^{1}$, Seiji Matsumoto$^{2}$, and Minoru Biyajima$^{2}$\\
{\small $^{1}$National Institute of Technology, Toba College, Toba 517-8501, Japan}\\
{\small $^{2}$Center for General Education, Shinshu University, Matsumoto 390-8621, Japan}}

\date{}
\maketitle

\begin{abstract}
The L3 Collaboration reported data on 2-jet and 3-jet Bose--Einstein correlations (BECs) with the results obtained through the $\tau$-model in 2011. In this study, we analyze these correlations using the conventional formula with the Gaussian long-range correlation (${\rm CF_I\times LRC_{(Gauss)}}$). The estimated ranges of interactions for 2-jet and 3-jet, $R_{\rm 2\mathchar`-jet}$ and $R_{\rm 3\mathchar`-jet}$, are almost the same magnitude as those by $\tau$-model: $R_{\rm 2\mathchar`-jet}=0.83\pm 0.05$ (stat) fm and $R_{\rm 3\mathchar`-jet}=1.09\pm 0.04$ (stat) fm. The anticorrelation in BEC (less than 1.0) observed by the L3 Collaboration is related to the partially negative density profile ($\mbox{\Large $\rho$}_{\rm \tau, BE}(\xi)$ in $\xi$ space) in the $\tau$-model and the ${\rm LRC_{(Gauss)}}=1/(1+\alpha e^{-\beta Q^2})$ in ${\rm CF_I\times LRC_{(Gauss)}}$, respectively. Remarkably, the probability $P_{\tau}(\xi)$ calculated from the Levy canonical form exhibits partially negative behavior. 
\end{abstract}

\section{\label{sec1}Introduction}
The L3 Collaboration reported their data on 2-jet and 3-jet Bose--Einstein correlation (BEC) in $e^+e^-$ collisions at Z$^0$-pole in 2011~\cite{L3:2011kzb}. The BEC is described in terms of the four momentum transfer $Q=\sqrt{-(p_1-p_2)^2}$,
\begin{eqnarray}
R_2 = \frac{\rho_2(p_1,\,p_2)}{\rho_0(p_1,\,p_2)}= \frac{\rho_2(Q)}{\rho_0(Q)}
\label{eq1}
\end{eqnarray}
where the numerator is the two-particle (mainly pions) distribution including the BE effect and the denominator is the two-particle distribution without the BE effect which is the mixture procedure. That means the $\rho_0(Q)$ distribution made from different events. 

Furthermore, introducing various corrections, L3 Collaboration adopted the following ratio expressed by Eq.~(\ref{eq1}),
\begin{eqnarray}
R_2^{\rm (L3)} = \frac{R_{\rm 2\,data}\cdot R_{\rm 2\,gen}}{R_{\rm 2\,det}\cdot R_{\rm 2\,gen\mathchar`-noBE}}.
\label{eq2}
\end{eqnarray}
In Eq.~(\ref{eq2}), the suffixes data, gen, det, and gen-noBE mean the following ensembles, respectively:
\begin{description}
  \item[1) data:] the data sample.
  \item[2) gen:] a generator-level Monte Carlo (MC) sample.
  \item[3) det:] the same MC sample passed through detector simulation and subject to the same selection procedure as the data.
  \item[and 4) gen-noBE:] a generator-level sample of a MC generated without BEC simulation.
\end{description}

Finally, $R_2^{\rm (L3)}$ data are analyzed by the following $\tau$-model based on the Levy canonical form to explain the anticorrelation (BEC: less than 1.0) observed in the interval $0.5\ {\rm GeV}\ Q<1.5$ GeV:
\begin{eqnarray}
F_{\rm \tau}=\left[1+\lambda\cos\left((R_a Q)^{2\alpha_{\tau}}\right)\exp\left(-(RQ)^{2\alpha_{\tau}}\right)\right]\times {\rm LRC_{(linear)}},
\label{eq3}
\end{eqnarray}
where $R$ denotes the magnitude of the interaction region and $R_a$ does a constrain expressed by $R_a^{2\alpha_{\tau}}=\tan(\alpha_{\tau}\pi/2)R^{2\alpha_{\tau}}$. The long-range correlation (${\rm LRC_{(linear)}}$) is expressed as ${\rm LRC_{(linear)}}=C(1+\delta Q)$. The parameter $\lambda$ denotes the degree of coherence, and $\alpha_{\tau}$ represents the characteristic index introduced in the Levy canonical form in the stochastic theory~\cite{CMS:2019fur,CMS:2011nlc}.\\

In contrast, OPAL~\cite{Acton:1991xb}, DELPHI~\cite{DELPHI:1992axf} and ALEPH Collaborations~\cite{ALEPH:1991loh} reported their data on BEC at Z$^0$-pole in 1991 and 1992, using the double ratio (DR) described with the single ratios $C_2^{\rm data}(Q)$ and $C_2^{\rm MC}(Q)$, 
\begin{eqnarray}
{\rm DR} =\frac{C_2^{\rm data}(Q)}{C_2^{\rm MC}(Q)} = \frac{N^{(2+:2-)}(Q)/N^{(+-)}(Q)}{N_{\rm MC}^{(2+:2-)}(Q)/N_{\rm MC}^{(+-)}(Q)},
\label{eq4}
\end{eqnarray}
where $N$ denotes the number of events. The suffixes $(2+:2-)$ and $(+-)$ indicate charge combinations.

It should be remarked that the reported data on BEC at Z$^0$-pole~\cite{Acton:1991xb,DELPHI:1992axf,ALEPH:1991loh} were analyzed by the conventional formula with ${\rm LRC_{(OPAL)}}=C(1+\delta Q+\varepsilon Q^2)$ and ${\rm LRC_{(linear)}}$,
\begin{eqnarray}
{\rm CF_I} = (1+\lambda\ E_{\rm BE})\times {\rm LRC},
\label{eq5}
\end{eqnarray}
where $E_{\rm BE}$ is the exchange function due to Bose--Einstein statistics for the identical pions. 

Thus, to compare fit parameters by L3 Collaboration and those by OPAL, DELPHI and ALEPH Collaborations, we have to analyze data on $R_2^{\rm (L3)}$ by Eq.~(\ref{eq5}). This is one of the aims of the present study.\\

\begin{figure}[H]
  \centering
  \includegraphics[width=0.48\columnwidth]{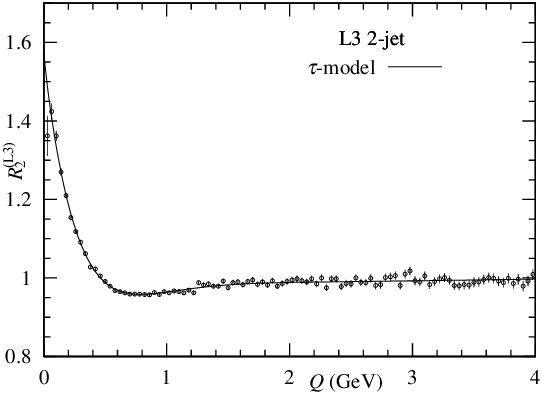}
  \includegraphics[width=0.48\columnwidth]{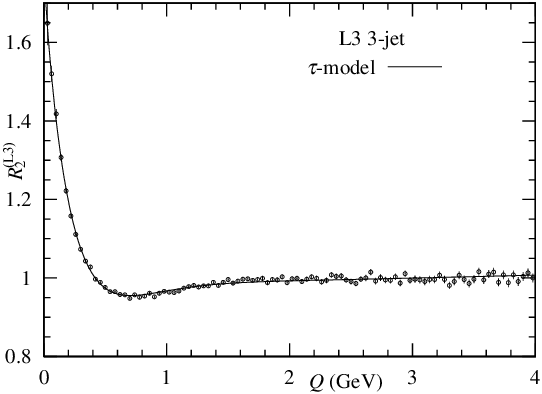}
  \caption{\label{fig1}Analysis of data on BEC measured as $R_2^{\rm (L3)}$ by Eq.~(\ref{eq3}).}
\end{figure}

First, to know the role of ${\rm LRC_{(linear)}}$, we examine data measured as $R_2^{\rm (L3)}$ by L3 Collaboration~\cite{L3:2011kzb} using Eq.~(\ref{eq3}). Table~\ref{tab1} and Fig.~\ref{fig1} present the results estimated using Eq.~(\ref{eq3}). As seen in Table~\ref{tab1}, the role of ${\rm LRC_{(linear)}}$ is considerably important because $\chi^2$/dof's are improved. In other words, the term of $\cos\left((R_a Q)^{2\alpha_{\tau}}\right)$ cannot sufficiently explain the anticorrelation. To show the role of $R_a^{2\alpha_{\tau}}=\tan(\alpha_{\tau}\pi/2)R^{2\alpha_{\tau}}$, we present our calculation with the constraint $R_a^{2\alpha_{\tau}}\times {\rm LRC_{(linear)}}$ in \S~4.

\begin{table}[H]
\centering
\caption{\label{tab1}Fit parameters of data on 2-jet and 3-jet events by Eq.~(\ref{eq3}). $R_a^{\alpha_{\tau}}=\tan(\alpha_{\tau}\pi/4)R^{\alpha_{\tau}}$ is calculated in the exact form. In the upper column, $\delta=0$ is used.}
\vspace{1mm}
\begin{tabular}{ccccccc}
\hline
event 
& $R$ (fm) 
& $\lambda$ 
& $C$ 
& $\alpha_{\tau}$ 
& $\delta$ (GeV$^{-1}$) 
& $\chi^2$/dof\\
\hline
2-jet
& $0.75\pm 0.03$
& $0.58\pm 0.02$
& $0.986\pm 0.001$
& $0.45\pm 0.01$
& ---
&  119.7/96\\
3-jet
& $0.93\pm 0.03$
& $0.78\pm 0.03$
& $0.989\pm 0.001$
& $0.44\pm 0.01$
& ---
&  226.8/96\\
\hline
2-jet
& $0.78\pm 0.04$
& $0.61\pm 0.03$
& $0.979\pm 0.002$
& $0.44\pm 0.01$
& $(4.6\pm 0.9)\times 10^{-3}$
& 94.6/95\\
3-jet
& $0.99\pm 0.04$
& $0.85\pm 0.04$
& $0.977\pm 0.001$
& $0.41\pm 0.01$
& $(7.7\pm 0.7)\times 10^{-3}$
& 113/95\\
\hline
2-jet
& $0.80\pm 0.04$
& $0.64\pm 0.04$
& $0.972\pm 0.005$
& $0.43\pm 0.01$
& $(1.40\pm 0.57)\times 10^{-2}$
& 91.5/94\\
&&&& \multicolumn{2}{c}{$\varepsilon=(-2.22\pm 1.37)\times 10^{-1}$ GeV$^{-2}$}\\

3-jet
& $1.06\pm 0.05$
& $0.93\pm 0.05$
& $0.963\pm 0.004$
& $0.40\pm 0.01$
& $(2.59\pm 0.44)\times 10^{-2}$
& 92.6/94\\
&&&& \multicolumn{2}{c}{$\varepsilon=(-4.40\pm 1.07)\times 10^{-1}$ GeV$^{-2}$}\\
\hline
\end{tabular}
\end{table}

Second, to know the role of ${\rm CF_I}$, we analyze the 2-jet and 3-jet BECs using Eq.~(\ref{eq5}) and the conventional formula ${\rm CF_I}$ with ${\rm LRC_{(Gauss)}}$ which is expressed as follows:
\begin{eqnarray}
{\rm CF_{I,\,Gauss}} = (1+\lambda\ E_{\rm BE})\times LRC_{\rm (Gauss)},
\label{eq6}\\
\mbox{where\quad}
{\rm LRC_{(Gauss)}} = \frac{C}{1+\alpha\exp(-\beta Q^2)}.
\nonumber
\end{eqnarray}
where ${\rm LRC_{(Gauss)}}$ was derived by us in Ref.~\cite{Mizoguchi:2023zfu} via analysis of CMS MC events at 13 TeV in LHC reported in Ref.~\cite{CMS:2019fur}. On the contrary, it should be noted that L3 Collaboration did not report their MC data. Thus we assume ${\rm LRC_{(Gauss)}}$ in the present study. See Ref.~\cite{Mizoguchi:2021slz}, for the sake of reference. Nevertheless, it should be stressed that the Gauss distribution is stable distribution in the sense of the Levy canonical form. See Appendix~\ref{secB}.

We consider the following four functions based on our previous study~\cite{Shimoda:1992gb}: 
\begin{eqnarray*}
\left\{
\begin{array}{l}
\qquad E_{\rm BE}=\exp(-(RQ)^2) \mbox{ (Gaussian distribution)},
\medskip\\
\qquad E_{\rm BE}=\exp(-RQ) \mbox{ (Exponential function)},
\medskip\\
\mbox{and\ }
E_{\rm BE}=\frac 1{(1+(RQ)^2)^s} \mbox{ (Inverse power low: $s=2$ and $s=5/2$)}. 
\end{array}
\right.
\end{eqnarray*}
The remainder of this paper is organized as follows. In Section~\ref{sec2}, we analyze the data for 2-jet and 3-jet using Eqs.~(\ref{eq5}) and (\ref{eq6}). In Section~\ref{sec3}, we analyze the same data by $\tau$-model$\times {\rm LRC_{(Gauss)}}$. The second aim of this paper is to examine connections in $\xi$-space between the quantum optics (QO) and the $\tau$-model in the stochastic approach. In Section~\ref{sec4}, introducing the Fourier transformation in the 4-dimensional Euclidean space-time (see Eq.~(\ref{eq11})), we can study the profiles of the exchange function $E_{\rm BE}$ of 2-jet and 3-jet in $\xi$-space, where $\xi=\sqrt{(x_1-x_2)^2+(y_1-y_2)^2+(z_1-z_2)^2+(ct_1-ct_2)^2}$. Finally, Section~\ref{sec5} presents the concluding remarks. 

In Appendix~\ref{secA}, data on $N_{\rm MC}^{(2+:2-)}/N_{\rm MC}^{(+-)}$ at $Z^0$-pole by DELPHI Collaboration~\cite{DELPHI:1992axf} are presented, because of no-information by L3 Collaboration. Therein we analyze DELPHI data by ${\rm LRC_{(linear)}}$ and ${\rm LRC_{(Gauss)}}$. In Appendix~\ref{secB}, the Levy canonical form and several formulations calculated using the inverse Fourier transformation are presented. In Appendix~\ref{secC}, first, the systematic error and uncertainties are investigated using of BEC at Z$^0$-pole by OPAL Collaboration~\cite{Acton:1991xb}, where  ${\rm CF_I(Gauss)\times LRC_{(OPAL)}}$ and ${\rm CF_I(Gauss)\times LRC_{(Gauss)}}$ are used. Second, the same quantities on BEC by L3 Collaboration are concisely mentioned.

\section{\label{sec2}Analysis of data on 2-jet and 3-jet using ${\rm CF_I}$ with ${\rm LRC_{(linear)}}$ and ${\rm LRC_{(Gauss)}}$}

\subsection*{I: Application of Eqs.~(\ref{eq5}) and (\ref{eq6})}
Using Eqs.~(\ref{eq5}) and (\ref{eq6}), we analyze data on the 2-jet and 3-jet with ${\rm LRC_{(linear)}}$ and ${\rm LRC_{(Gauss)}}$. Our results are presented in Table~\ref{tab2}.

\begin{figure}[H]
  \centering
  \includegraphics[width=0.48\columnwidth]{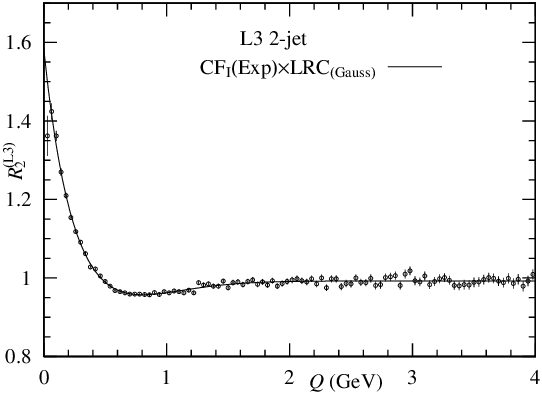}
  \includegraphics[width=0.48\columnwidth]{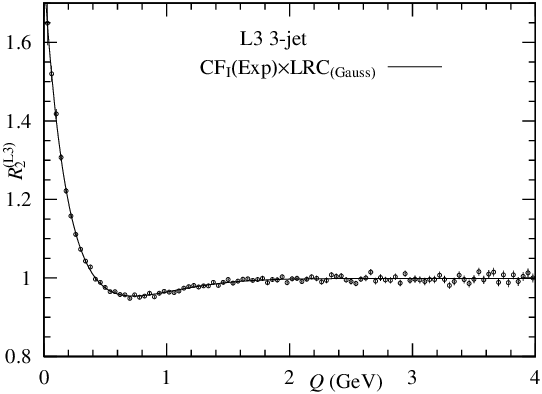}
  \caption{\label{fig2}Analysis of data using Eqs.~(\ref{eq5}) and (\ref{eq6})}
\end{figure}

\begin{table}[H]
\centering
\caption{\label{tab2}Fit parameters of data on 2-jet and 3-jet events using ${\rm CF_I}$ with ${\rm LRC_{(linear)}}$ and ${\rm LRC_{(Gauss)}}$. Note that $p$-values for ${\rm CF_I(Exp)\times LRC_{(Gauss)}}$ are 62.6 \% (2-jet) and 78.5 \% (3-jet), respectively.}
\vspace{1mm}

\begin{tabular}{ccccccc}
\hline
\multicolumn{2}{l}{${\rm LRC_{(linear)}}$}\\
$E_{\rm BE}$ 
& $R$ (fm) 
& $\lambda$ 
& $C$ 
& \multicolumn{2}{c}{$\delta$ (GeV$^{-1}$)} 
& $\chi^2$/dof\\
\hline
2-jet\\
Gauss
& $ 0.68\pm 0.01$
& $ 0.41\pm 0.01$
& $0.956\pm 0.002$
& \multicolumn{2}{c}{$(1.34\pm 0.09)\times 10^{-2}$}
& 247/96\\

Exp.
& $ 1.18\pm 0.02$
& $ 0.80\pm 0.02$
& $0.947\pm 0.002$
& \multicolumn{2}{c}{$(1.75\pm 0.10)\times 10^{-2}$}
& 255/96\\

IP2.0
& $ 0.64\pm 0.01$
& $ 0.52\pm 0.01$
& $0.946\pm 0.002$
& \multicolumn{2}{c}{$(1.79\pm 0.10)\times 10^{-2}$}
& 247/96\\

IP2.5
& $ 0.54\pm 0.01$
& $ 0.49\pm 0.01$
& $0.949\pm 0.002$
& \multicolumn{2}{c}{$(1.67\pm 0.10)\times 10^{-2}$}
& 231/96\\

\hline
3-jet\\
Gauss
& $ 0.79\pm 0.01$
& $ 0.51\pm 0.01$
& $0.953\pm 0.001$
& \multicolumn{2}{c}{$(1.74\pm 0.07)\times 10^{-2}$}
& 455/96\\

Exp.
& $ 1.44\pm 0.02$
& $ 1.06\pm 0.02$
& $0.945\pm 0.001$
& \multicolumn{2}{c}{$(2.10\pm 0.08)\times 10^{-2}$}
& 438/96\\

IP2.0
& $ 0.77\pm 0.01$
& $ 0.68\pm 0.01$
& $0.943\pm 0.001$
& \multicolumn{2}{c}{$(2.19\pm 0.08)\times 10^{-2}$}
& 472/96\\

IP2.5
& $ 0.65\pm 0.01$
& $ 0.64\pm 0.01$
& $0.946\pm 0.001$
& \multicolumn{2}{c}{$(2.07\pm 0.08)\times 10^{-2}$}
& 436/96\\

\hline

\hline
\multicolumn{2}{l}{${\rm LRC_{(Gauss)}}$}\\
$E_{\rm BE}$ 
& $R$ (fm) 
& $\lambda$ 
& $C$ 
& $\alpha$ 
& $\beta$ (GeV$^{-2}$) 
& $\chi^2$/dof\\
\hline
2-jet\\
Gauss
& $ 0.65\pm 0.01 $
& $ 0.40\pm 0.01 $
& $0.995\pm 0.002$
& $0.046\pm 0.003$
& $ 0.49\pm 0.07 $
&  199/95\\

Exp.
& $ 0.83\pm 0.05 $
& $ 0.82\pm 0.03 $
& $0.992\pm 0.001$
& $0.137\pm 0.024$
& $ 1.15\pm 0.12 $
&  90.0/95\\

IP2
& $ 0.55\pm 0.01 $
& $ 0.52\pm 0.01 $
& $0.993\pm 0.001$
& $0.802\pm 0.006$
& $ 0.77\pm 0.08 $
&  118/95\\

IP2.5
& $ 0.48\pm 0.01 $
& $ 0.49\pm 0.01 $
& $0.994\pm 0.002$
& $0.701\pm 0.005$
& $0.725\pm 0.08 $
& 126/95\\
\hline
3-jet\\
Gauss
& $ 0.75\pm 0.01 $
& $ 0.50\pm 0.01 $
& $1.001\pm 0.001$
& $0.058\pm 0.002$
& $ 0.57\pm 0.05 $
&  310/95\\

Exp.
& $ 1.09\pm 0.04 $
& $ 0.99\pm 0.02 $
& $0.999\pm 0.001$
& $0.115\pm 0.009$
& $ 1.06\pm 0.08 $
&  83.9/95\\

IP2.0
& $ 0.65\pm 0.01 $
& $ 0.65\pm 0.01 $
& $1.000\pm 0.001$
& $0.093\pm 0.004$
& $ 0.84\pm 0.06 $
&  140/95\\

IP2.5
& $ 0.56\pm 0.01 $
& $ 0.62\pm 0.01 $
& $1.000\pm 0.001$
& $0.083\pm 0.004$
& $ 0.79\pm 0.06 $
&  157/95\\
\hline

\end{tabular}
\end{table}

As shown in Tables~\ref{tab1} and ~\ref{tab2}, our results from Eq.~(\ref{eq6}) and ${\rm LRC_{(Gauss)}}$ are compatible with those presented in Table~\ref{tab1}. However, the degree of coherence $\lambda$'s presented in Table~\ref{tab1} are somewhat smaller than those presented in Table~\ref{tab2}. 

Moreover, we would like to add the following: For 2-jet, as we assume the Levy distribution ($\exp(-(RQ)^{\alpha_{\rm L}})$) times ${\rm LRC_{(Gauss)}}$, we obtain the similar figures as ${\rm CF_I(Exp)\times LRC_{(Gauss)}}$, because of the Levy index $\alpha_{\rm L}=1.027\pm 0.107$ ($\chi^2/{\rm dof}=89.9/94$). For 3-jet, there is no coincidence with the result by ${\rm CF_I(Exp)\times LRC_{(Gauss)}}$, because of $\alpha_{\rm L}=0.865\pm 0.087$ and $\lambda = 1.194\pm 0.161$.

\subsection*{II: Analysis of 2-jet and 3-jet events using the quantum optics  approach}
As shown in Table~\ref{tab2}, the degree of coherence $\lambda=0.82\pm 0.03$ in ${\rm CF_I(exp)\times LRC_{(Gauss)}}$ is observed. To consider its meaning, we use the following formula with QO~\cite{Biyajima:1979ak,Biyajima:1990ku,Kozlov:2007dv}.
\begin{eqnarray}
F_{\rm QO}=(1+p^2e^{-RQ}+2p(1-p)e^{-RQ/2})\times{\rm LRC_{(Gauss)}},
\label{eq7}
\end{eqnarray}
where $p=A/\langle n\rangle_{\rm tot}$ represents the ratio of the chaotic component to the total multiplicity. $1-p=\langle n\rangle_{\rm coherent}/\langle n\rangle_{\rm tot}$ denotes the ratio of the coherent component to the total one. For the 2-jet event, as shown in Table~\ref{tab3}, $1-p\cong 0.3$, the coherent component is approximately 30\%. Moreover, $R\cong 1.06\pm 0.10$ fm is larger than those in Table~\ref{tab2}. This is attributed to the third term $2p(1-p)e^{-RQ/2}$.

\begin{table}[H]
\centering
\caption{\label{tab3}Fit parameters of data on 2-jet and 3-jet events using Eq.~(\ref{eq7})}
\vspace{1mm}
\begin{tabular}{ccccccc}
\hline
event 
& $R$ (fm) 
& $p$ 
& $C$ 
& $\delta$ (GeV$^{-1}$)
& $\varepsilon$ (GeV$^{-2}$) 
& $\chi^2$/dof\\
\hline
2-jet
& $1.892\pm 0.03$
& $0.612\pm 0.03$
& $0.943\pm 0.002$
& $(1.91\pm   0.10)\times 10^{-2}$
& ---
&  314/96\\

3-jet
& $1.392\pm 0.12$
& $1.0$
& $0.945\pm 0.001$
& $(2.13\pm  0.08)\times 10^{-2}$
& ---
&  447/96\\

\hline
2-jet
& $1.42\pm 0.05$
& $0.72\pm 0.04$
& $0.878\pm 0.007$
& $(10.1\pm 0.8)\times 10^{-2}$
& $(-1.8\pm 0.2)\times 10^{-2}$
&  156/95\\

3-jet
& $1.18\pm 0.02$
& $1.0$
& $0.896\pm 0.004$
& $(8.8\pm 0.5)\times 10^{-2}$
& $(-1.6\pm 0.1)\times 10^{-2}$
&  167/95\\

\hline
event 
& $R$ (fm) 
& $p$ 
& $C$ 
& $\alpha$ 
& $\beta$ (GeV$^{-2}$) 
& $\chi^2$/dof\\
\hline
2-jet
& $1.06\pm 0.10$
& $0.71\pm 0.04$
& $0.992\pm 0.001$
& $0.19\pm 0.02$
& $1.12\pm 0.05$
& 90.4/95\\
3-jet
& $1.11\pm 0.04$
& $0.98\pm 0.02$
& $0.999\pm 0.001$
& $0.12\pm 0.01$
& $1.07\pm 0.07$
& 83.6/95\\
\hline
\end{tabular}
\end{table}

For the 3-jet event, in Eq.~(\ref{eq7}), we obtain $p=0.98\pm 0.02$, meaning that $p=A/\langle n\rangle_{\rm tot}\cong 1.0$. There is no contribution from the coherent component, because $1-p\cong 0.0$. We conclude that the data on the 3-jet events are almost chaotic~\cite{HanburyBrown:1956bqd}. See also Ref.~\cite{Goldhaber:1960sf} for comparison.

\section{\label{sec3}Analysis of 2-jet and 3-jet events using the $\tau$-model including ${\rm LRC_{(Gauss)}}$}
To examine the role of ${\rm LRC_{(Gauss)}}$ in Eq.~(\ref{eq3}), we propose the following formula, 
\begin{eqnarray}
F_{\rm \tau,\,Gauss}=\left[1+\lambda\cos\left((R_a Q)^{2\alpha_{\tau}}\right)\exp\left(-(RQ)^{2\alpha_{\tau}}\right)\right]\times {\rm LRC_{(Gauss)}},
\label{eq8}
\end{eqnarray}
Using of sets of six random variables and the CERN MINUIT program, we can estimate the fit parameters shown in Table~\ref{tab4}.

\begin{table}[H]
\centering
\caption{\label{tab4}Fit parameters of L3 Collaboration events, i.e., $R_2^{\rm (L3)}$ by using Eq.~(\ref{eq8}).}
\vspace{1mm}
\begin{tabular}{cccccccc}
\hline
event 
& $R$ (fm) 
& $\lambda$ 
& $C$ 
& $\alpha_{\tau}$ 
& $\alpha$
& $\beta$ (GeV$^{-2}$) 
& $\chi^2$/dof\\
\hline
2-jet
& $0.82\pm 0.05$
& $0.67\pm 0.12$
& $0.992\pm 0.001$
& $0.42\pm 0.03$
& $0.03\pm 0.05$
& $1.11\pm 1.90$
&  91.2/94\\

3-jet
& $1.13\pm 0.06$
& $1.02\pm 0.08$
& $0.999\pm 0.001$
& $0.37 \pm 0.01$
& $0.05\pm  0.02$
& $1.19\pm  0.37$
&  83.5/94\\
\hline
\end{tabular}
\end{table}

The differences concerning $\chi^2$/dof's in Tables~\ref{tab1} and \ref{tab4} are obvious. It is worthwhile to stress that the degree of coherence $\lambda$s is larger than those by ${\rm LRC_{(linear)}}$ in Table~\ref{tab1}. Moreover, the magnitudes of $R$s are almost the same as those by the exponential function in Table~\ref{tab2}. For 3-jet, $R=1.11\pm 0.04$ (stat) fm estimated using Eq.~(\ref{eq7}) in Table~\ref{tab3} is compatible with $R=1.13\pm 0.06$ (stat) fm obtained  using Eq.~(\ref{eq8}) in Table~\ref{tab4}.

\section{\label{sec4}Mechanism of anticorrelation and profiles of the source functions of 2-jet and 3-jet BECs in the $\xi$-space}
%
\subsection*{I: Explanation of the anticorrelation}
The anticorrelation in the 2-jet and 3-jet events, i.e., $R_2^{\rm (L3)}$ is observed in the region $0.5\le Q\le 1.5$ GeV. These values in BEC are smaller than 1.0. These behaviors can be explained in two ways. \\

In the conventional formula with the function, ${\rm CF_I(Exp)\times LRC_{(Gauss)}}$, the anticorrelation is explained by the Gaussian function with ${\rm LRC_{(Gauss)}-1} = \sum_{k=1}^{\infty}(-\alpha)^k\exp(-k\beta Q^2)$. Second, the cross term which is a product of the exponential function and  $({\rm LRC_{(Gauss)}-1}$): $F_{\rm cross}(Q) = e^{-RQ}\sum_{k=1}^{\infty}(-\alpha)^ke^{-k\beta Q^2}$. For reference, we present data on $N_{\rm MC}^{(2+:2-)}/N_{\rm MC}^{(+-)}$ at $Z^0$-ploe by DELPHI Collaboration~\cite{DELPHI:1992axf} in Appendix~\ref{secA} because we have no-information on them ($R_{\rm 2\,data}$, $R_{\rm 2\,gen\mathchar`-noBE}$ and $(R_{\rm 2\,gen}/R_{\rm 2\,det})$) by L3 Collaboration at present.\\

\begin{figure}[H]
  \centering
  \includegraphics[width=0.48\columnwidth]{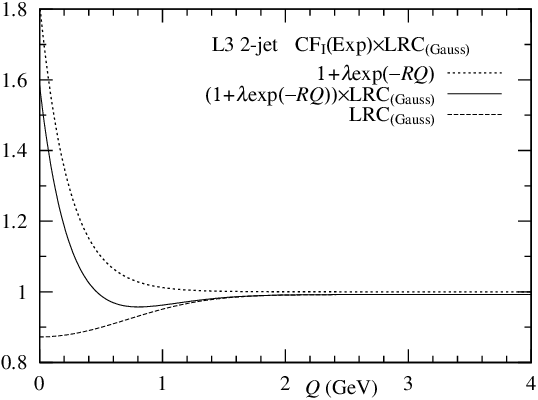}\\
  \includegraphics[width=0.48\columnwidth]{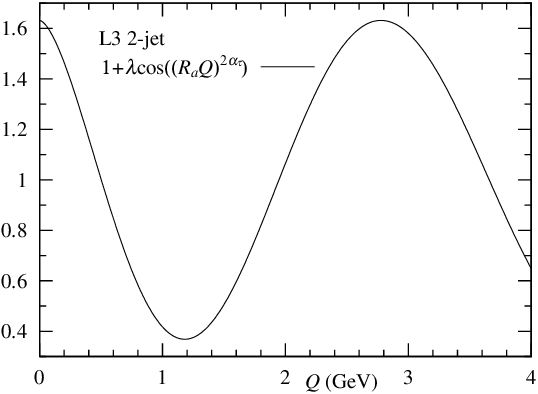}
  \includegraphics[width=0.48\columnwidth]{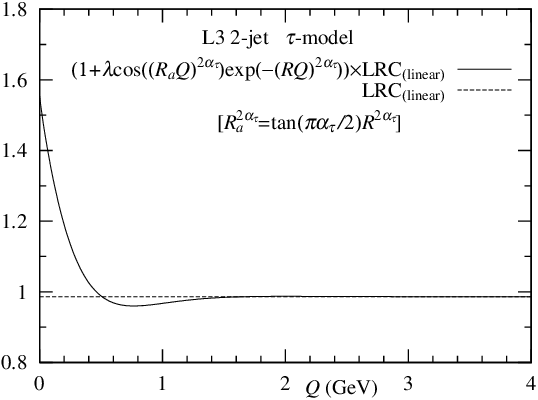}
  \caption{\label{fig3}Mechanism of ``anticorrelation'' in ${\rm CF_I\times LRC_{(Gauss)}}$, i.e., Eq.~(\ref{eq5}) and the $\tau$-model, i.e., Eq.~(\ref{eq3}).}
\end{figure}

Alternatively, in the $\tau$-model, the anticorrelation is described by the cosine $\left[(R_a Q)^{\alpha_{\tau}}\right]$ in the exchange function. Indeed, this mechanism is shown in Fig.~\ref{fig3}. The behavior of $\cos((R_aQ)^{\alpha_{\tau}})$ is oscillating. By multiplying $\exp(-(RQ)^{\alpha_{\tau}})$, the anticorrelation is created. Moreover, when ${\rm LRC_{(linear)}}$ was replaced by ${\rm LRC_{(Gauss)}}$ in $\tau$-model, we observe more significant anticorrelation. See Fig.~\ref{fig4}.

\subsection*{II: Source functions of BEC in the $\xi$-space}
Using the Fourier transform of $E_{\rm BE}$'s and ${\rm LRC_{(Gauss)}}$, we calculate the profiles of ${\rm CF_I\times LRC_{(Gauss)}}$  and the $\tau$-model in the $\xi$-space, which are expressed as the density $\rho(\xi)$s.

We use the following expansion to study the profile of ${\rm CF_I\times LRC_{(Gauss)}}$ in the $\xi$-space. Before concrete computation, we add the physical meaning. $\mbox{\Large $\rho$}_{\rm Exp}(\xi)$ is interpreted as $N=4$ dimensional Lorentz distribution because $(N+1)/2=(4+1)/2=5/2$ (see Ref.~\cite{Uchaikin:1999}~\footnote{The chapter 7, ``multivariate stable laws'' is useful for the calculation of the general systematical distributions ($N>1$). The results in Ref.~\cite{Shimoda:1992gb} are consistent with those in Ref.~\cite{Uchaikin:1999} for $N=4$. See also Ref.~\cite{Zolotarev:1981aa}}). $\mbox{\Large $\rho$}_{\rm Gauss}(\xi)$ is connected to $\exp(-\beta Q^2)$ in the ${\rm LRC_{(Gauss)}}$~\cite{Shimoda:1992gb}. 
\begin{eqnarray}
\left\{
\begin{array}{l}
\mbox{\Large $\rho$}_{\rm Exp}(\xi) = \dfrac 3{4\pi^2R^4}\dfrac{1}{(1+(\xi/R)^2)^{5/2}},
\medskip\\
S_{\rm BE}(\xi) = 2\pi^2\xi^3\mbox{\Large $\rho$}_{\rm Exp}(\xi).
\end{array}
\right.
\label{eq9}
\end{eqnarray}
The source function of the ${\rm LRC_{(Gauss)}}$ is calculated as follows:
\begin{eqnarray}
\left\{
\begin{array}{l}
\mbox{\Large $\rho$}_{\rm Gauss}(\xi) = \dfrac{1}{16\pi^2R^4}\exp\left(-\dfrac{\xi^2}{4R^2}\right),
\medskip\\
S_{\rm BG}(\xi) = 2\pi^2\xi^3\displaystyle{\sum_{k=1}^{\infty}}(-\alpha)^k\mbox{\Large $\rho$}_{\rm Gauss}(\xi, R=\sqrt{k\beta}).
\end{array}
\right.
\label{eq10}
\end{eqnarray}
The contribution from the cross term, $e^{-RQ}\times{\rm LRC_{(Gauss)}}$, is computed by the inverse Fourier transformation ($N=4$) and the numerical integration. 

\begin{figure}[H]
  \centering
  \includegraphics[width=0.48\columnwidth]{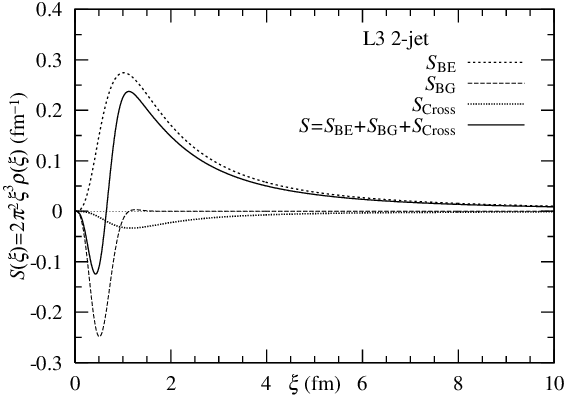}
  \includegraphics[width=0.48\columnwidth]{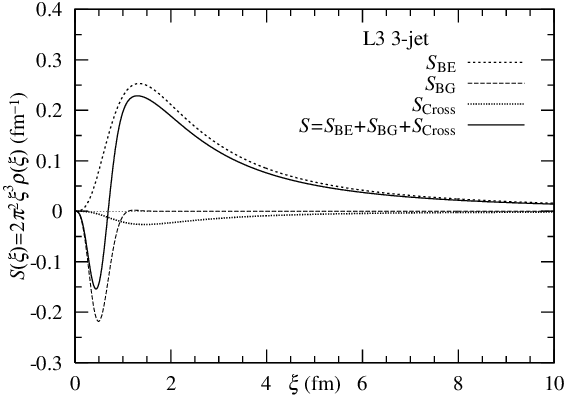}\\
  \includegraphics[width=0.48\columnwidth]{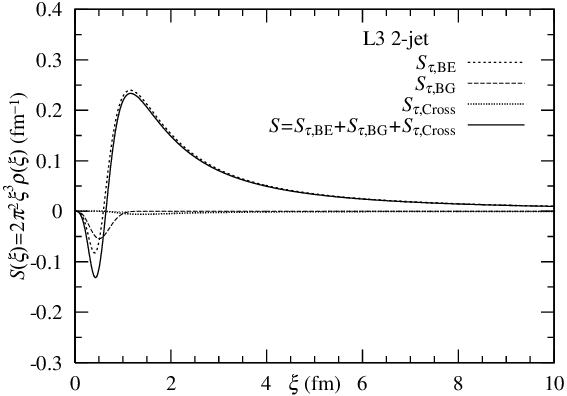}
  \includegraphics[width=0.48\columnwidth]{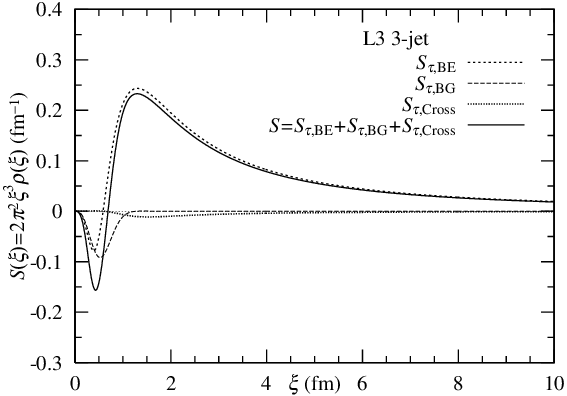}\\
  \caption{\label{fig4}Source functions of ${\rm CF_I(Exp)\times LRC_{(Gauss)}}-1$ and the $\tau$-model$\times {\rm LRC_{(Gauss)}}-1$ in $\xi$-space. The source functions of $({\rm LRC_{(Gauss)}-1})$ expressed by $S_{\rm BG}(\xi)$ are added in the upper figures. The same calculations on $S_{\tau,{\rm BG}}$ are made in lower ones. Estimated values in Tables~\ref{tab2} and \ref{tab4} are used in computations.}
\end{figure}

For the stochastic density of the exchange function $E_{\rm BE}$ in the $\tau$-model, we first employ the following numerical calculation:
\begin{eqnarray}
\left\{
\begin{array}{l}
E_{\rm BE}^{(\tau)}=\exp(-aQ_{\xi}^{2\alpha_{\tau}})\cos(bQ_{\xi}^{2\alpha_{\tau}}),
\medskip\\
\mbox{\Large $\rho$}_{\rm \tau,BE}(\xi) = \dfrac 1{(2\pi)^2\xi} \dint Q_{\xi}^2\, E_{\rm BE}^{(\tau)}\, J_1(Q_{\xi}\xi)dQ_{\xi}
\end{array}
\right.
\label{eq11}
\end{eqnarray}
where $J_1(Q_{\xi}\xi)$ is the Bessel function~\cite{Shimoda:1992gb}. Note that $Q_{\xi}\!=\!\sqrt{(p_{1x}\!-\!p_{2x})^2\!+\!(p_{1y}\!-\!p_{2y})^2\!+\!(p_{1z}\!-\!p_{2z})^2\!+\!(E_{1}\!-\!E_{2})^2}$ in Eq.~(\ref{eq11}). The Wick rotation is necessary~\cite{Wick:1954eu}. Fig.~\ref{fig4} shows the numerical calculation of Eq.~(\ref{eq11}). 

Second, concerning with the analytic expansion of Eq.~(\ref{eq11}), we obtain the following formula,
\begin{eqnarray}
&&\mbox{Series expansion of } \mbox{\Large $\rho$}_{\rm \tau,BE}(\xi)\nonumber\\
&&= \frac 1{(2\pi)^2\xi} \frac 1{\alpha_{\tau}}\sum_{k=0}^{\infty}\left(\frac{\xi}2\right)^{2k+1}\dfrac{(-1)^k\Gamma\left(\frac{4+2k}{\alpha_{\tau}}\right)}{\Gamma(k+1)\Gamma(k+2)}\frac 1{(a^2+b^2)^{(4+2k)/(2\alpha_{\tau})}}
\cos\left(\frac{4+2k}{\alpha_{\tau}}\arctan\left(\frac ba\right)\right)
\label{eq12}
\end{eqnarray}
where $a=0.8293$, $b=0.7366$ and $2\alpha_{\tau}=0.8104$ (see Table~\ref{tab1}). However, it is regretful that we cannot obtain the full numerical values, because of the mathematical limit of serious expansion of the Bessel function. Note that $\xi=0.3$ fm is the applicable limit of Eq.~(\ref{eq12}). 

Third, we calculate profiles of Eqs~(\ref{eq6}) and (\ref{eq8}) in $\xi$-space. Fig.~\ref{fig4} shows the two types of behaviors for the 2-jet and 3-jet events. The dip structures at $\xi\sim 0.5$ fm are observed. The dip structures in the 3-jet (right) are somewhat deeper than those in the 2-jet (left).

It can be said that, the negative profile of the density $\mbox{\Large $\rho$}(\xi)$ in ${\rm CF_I\times LRC_{(Gauss)}}$ originates from the denominator. In other words, the Monte Carlo calculation in the denominator in L3 BEC at Z$^0$-pole implies the extent of the clustering effect expressed by the Gaussian distribution as well as in the $N_{\rm MC}^{(2+;2-)}$ for LHC CMS BEC at 13 TeV~\cite{CMS:2019fur,Mizoguchi:2023zfu}.

\begin{table}[H]
\centering
\caption{\label{tab5}Comparison of physical pictures of our approach (${\rm CF_I(Exp)\times LRC_{(Gauss)}}$) and the $\tau$-model. In the $\tau$-model, $R_{\rm 2\,data}$, $R_{\rm 2\,det}$ and $(R_{\rm 2\,gen}/R_{\rm 2\,gen\mathchar`-noBE})$ are not separated.}
\vspace{1mm}
\renewcommand{\arraystretch}{1.4}
\begin{tabular}{c|ll}
\hline
& $\xi$-space & Momentum ($Q$)-space\\

\hline

Our approach 
& The numerator:
& 
\\
(${\rm CF_I(Exp)}\quad$
& $\mbox{\Large $\rho$}_{\rm Lorentz}(\xi,N=4) = \dfrac 3{4\pi^2R^4}\dfrac{1}{(1+(\xi/R)^2)^{5/2}}$
& \quad $E_{\rm BE}=e^{-RQ}$.
\\
$\times{\rm LRC_{(Gauss)}}$)
& where $5/2=(N+1)/2$. $N=4$ means 
& 
\\
& 4-dimension.
& 
\\
& The denominator of LRC:
& \quad ${\rm LRC_{(Gauss)}} = \dfrac{C}{1+\alpha\exp(-\beta Q^2)}$
\\
& $\mbox{\Large $\rho$}_{\rm Gauss}(\xi, N=4) = \dfrac{1}{16\pi^2R^4}\exp\left(-\dfrac{\xi^2}{4R^2}\right)$.
& \quad \qquad\ $=C\displaystyle{\sum_{k=0}^{\infty}} (-\alpha)^k e^{-k\beta Q^2}$.
\\
& See Ref.~\cite{Shimoda:1992gb}.
& which is reflecting the clustering effect
\\
& 
& observed in $N_{\rm MC}^{(2+:2-)}$~\cite{Mizoguchi:2023zfu,CMS:2019fur}. See Ref.~\cite{DELPHI:1992axf}.
\\

\hline

$\tau$-model 
& The probability density and source function 
& Levy canonical form ($\zeta = \tan(\alpha_{\tau}\pi/2)$)
\\
& are calculated from $\tilde H_{4}(\omega)$ in the right column.
& \quad $\tilde H_{4}(RQ) = \exp\left[-\dfrac 12(RQ)^{2\alpha_{\tau}}(1-i\,\zeta)\right]$,
\\
& $\mbox{\Large $\rho$}_{\rm \tau, BE}(\xi, \alpha_{\tau}) = \dfrac 1{(2\pi)^2\xi} \dint Q_{\xi}^2\, \exp(-aQ_{\xi}^{2\alpha_{\tau}})$
& \quad $E_{\rm BE} = {\rm Re}\left[\tilde H_{4}^2(RQ)\right]$
\\
& \quad\qquad\qquad $\cdot\cos(bQ_{\xi}^{2\alpha_{\tau}})\, J_1(Q_{\xi}\xi)dQ_{\xi}$
& \qquad\quad $= \exp\left(-(RQ)^{2\alpha_{\tau}}\right)\cos\left((RQ)^{2\alpha_{\tau}}\zeta\right)$.
\\
& $S_{\rm \tau, BE}(\xi) =2\pi^2\xi^3\mbox{\Large $\rho$}_{\rm \tau, BE}(\xi, \alpha_{\tau})$
& See Appendix~\ref{secB}. ${\rm LRC}=C(1+\delta Q)$.
\\

\hline

\multicolumn{3}{l}{cf. The probability of the Levy canonical form ($N=1$ dimension) is expressed as}\\
\multicolumn{3}{c}{$P(x; \alpha_{\tau}, \theta)=\dfrac 1{\pi}\mbox{Re}\left[\dint dz\exp\left(-ixz-z^{\alpha_{\tau}}e^{i\frac{\pi}2\theta}\right)\right]$,}\\
\multicolumn{3}{l}{where Re[\ ] means the real part. The Lorentz (or Cauchy) distribution is obtained by }\\
\multicolumn{3}{l}{$P(x; 1, 0)=1/[\pi(1+x^2)]$. The Gaussian one is obtained by $P(x; 2, 0)=(1/\pi)e^{-x^2}$. The formulas}\\
\multicolumn{3}{l}{with $N=4$ are calculated by the following replacement in $P(x; \alpha_{\tau}, \theta)$ above: $ixz\to ixz\cos\phi$ and }\\
\multicolumn{3}{l}{$dz\to d^4z = \sin^2\phi d\phi\cdot 4\pi z^3dz$. $P_4(x; 1, 0)=1/[\pi(1+x^2)^{5/2}]$ and  $P_4(x; 2, 0)=\pi e^{-x^2/4}$ are obtained.}\\
\hline
\end{tabular}
\end{table}

Concerning the Levy canonical form and the inverse Fourier transformation ($N=1$ and $N=4$), several formulations are shown in Appendix~\ref{secB}.

\section{\label{sec5}Concluding remarks}

\paragraph{C1)}
From our analysis of L3 BEC by ${\rm CF_I(Exp)\times LRC_{(Gauss)}}$, we have known that the anticorrelation (observed in $0.5\ {\rm GeV}\ Q<1.5$ GeV) in related to the denominator of the DR. In the ${\rm CF_I(Exp)\times LRC_{(Gauss)}}$, $E_{\rm BE}=\exp(-RQ)$ cooperates with ${\rm LRC_{(Gauss)}}$ (see Fig.~\ref{fig4}). Provided that the MC events at Z$^0$-pole by L3 Collaboration were reported, we are able to know whether or not the statement mentioned above is correct.

\paragraph{C2)}
On the other hand, in the $\tau$-model, the anticorrelation is explained by the imaginary part in the Levy canonical form. Indeed, through the analytic form of the exchange function $\exp(-aQ^{2\alpha})\cos(bQ^{2\alpha})$ in the $\xi$-space, we understand that the anticorrelation in BEC is related to the negative behavior therein. 

\paragraph{C3)}
In the $\tau$-model, we obtain the analytic formula, i.e., the series expansion of $\mbox{\Large $\rho$}_{\rm \tau,BE}(\xi)$ (see Eq.~(\ref{eq12})). We can understand the partially negative behavior ($\xi<0.7$ fm) in the Levy canonical form in Fig.~\ref{fig4}.

\begin{figure}[H]
  \centering
  \includegraphics[width=0.6\columnwidth]{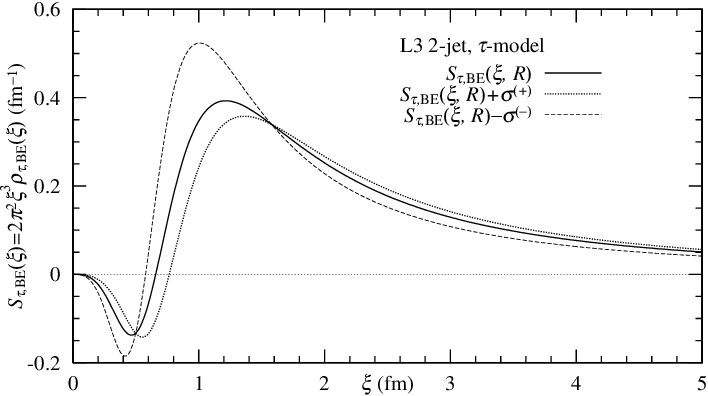}
  \caption{\label{fig5}Profile of source function $S_{\rm \tau, BE}(\xi)$ in the $\tau$-model. $\sigma^{(\pm)} = \partial S_{\rm \tau, BE}(\xi,\,R)/\partial R\cdot \delta R^{(\pm)}$ with $\delta R^{(+)}=\sqrt{(0.04(\rm stat))^2+(0.09(\rm sys))^2}$ and $\delta R^{(-)}=\sqrt{(0.04(\rm stat))^2+(0.16(\rm sys))^2}$. The values ($0.01_{-0.160}^{+0.09}$) are taken from Table 3 in Ref.~\cite{L3:2011kzb}, therein ${\rm LRC_{(linear)}}$ is used.}
\end{figure}

The profile of source function $S_{\rm \tau, BE}(\xi)$ in the $\tau$-model is described as a product of the phase factor $2\pi^2\xi^3$ and $\mbox{\Large $\rho$}_{\rm \tau, BE}(\xi)$. The source function $S_{\rm \tau, BE}(\xi)$ (with $R=0.78$ fm and $\alpha_{\tau} = 0.44$) is shown in Fig.~\ref{fig5}. We observe the partially negative behavior in the range $0<\xi<0.6$ fm, because of $\cos((RQ)^{2\alpha_{\tau}}\zeta)$.
%
%

\paragraph{C4)}
Two estimated values $R=0.83\pm 0.05$ (stat) fm (2-jet) and $R=1.09\pm 0.04$ (stat) fm (3-jet) by ${\rm CF_I(Exp)\times LRC_{(Gauss)}}$ (see Table~\ref{tab6}) are compared with $R=0.72\pm 0.04$ (stat) fm (2-jet) and $R=1.13\pm 0.06$ (stat) fm (3-jet) estimated by $\tau$-model$\times {\rm LRC_{(Gauss)}}$, respectively. It is remarked that the fitting parameters for the ${\rm CF_I(Exp)}$ and $\tau$-model with ${\rm LRC_{(Gauss)}}$ are almost coincident with each other.

\begin{table}[H]
\centering
\caption{\label{tab6}Comparison of fitting parameters for ${\rm CF_I(Exp)}$ and $\tau$-model. The $p$-values (\%) are shown in parentheses.}
\vspace{1mm}

\begin{tabular}{ccccccl}
\hline
2-jet & & & ${\rm LRC_{(linear)}}$ & & \multicolumn{2}{c}{${\rm LRC_{(Gauss)}}$ ($p$-value \%)}\\
\hline
${\rm CF_I(Exp)}$
& $\chi^2$/dof
&& 255/96
&&  90.0/95 & (62.6)\\

& $\lambda$ 
&& $ 0.80\pm 0.02$
&& $ 0.82\pm 0.03$\\

& $R$ (fm) 
&& $ 1.18\pm 0.02$
&& $ 0.83\pm 0.05$\\

\hline

$\tau$-model
& $\chi^2$/dof
&& 94.6/95
&& 91.2/94 & (56.3)\\

& $\lambda$ 
&& $0.61\pm 0.03$
&& $0.67\pm 0.12$\\

& $R$ (fm) 
&& $0.78\pm 0.04$
&& $0.82\pm 0.05$\\

\hline
\hline

3-jet & & & ${\rm LRC_{(linear)}}$ & & \multicolumn{2}{c}{${\rm LRC_{(Gauss)}}$ ($p$-value \%)}\\
\hline
${\rm CF_I(Exp)}$
& $\chi^2$/dof
&& 438/96
&&  83.9/95 & (78.5)\\

& $\lambda$ 
&& $ 1.06\pm 0.02$
&& $ 0.99\pm 0.02$\\

& $R$ (fm) 
&& $ 1.44\pm 0.02$
&& $ 1.09\pm 0.04$\\
\hline

$\tau$-model
& $\chi^2$/dof
&& 113/95
&& 83.5/94 & (77.3)\\

& $\lambda$ 
&& $0.85\pm 0.04$
&& $1.02\pm 0.08$\\

& $R$ (fm) 
&& $0.99\pm 0.04$
&& $1.13\pm 0.06$\\
\hline
\end{tabular}
\end{table}

From values in Table~\ref{tab4}, we can estimate the values of full width of half maximum (FWHM) for 2-jet and 3-jet. The estimated FWHMs and HWHM (fm)s are presented in Table~\ref{tab7}. It can be stressed that HWHM's in the $\tau$-model are coincident with $R$'s in Table~\ref{tab6}. In ${\rm CF_I(Exp)}$, the HWHM's increase about 10\% than $R$'s in Table~\ref{tab6}.

\begin{table}[H]
\renewcommand{\arraystretch}{1.3}
\centering
\caption{\label{tab7}Estimated FWHMs and HWHMs of the source functions in Fig.~\ref{fig4}.}
\vspace{1mm}

\begin{tabular}{ccccc}
\hline
&
& $\xi^{\rm H}-\xi^{\rm L}$ (fm) 
& FWHM (fm)
& HWHM (fm)\\
\hline
\lw{2-jet}
& ${\rm CF_I(Exp)}$
& $2.35-0.48$
& $1.87$
& $0.94$\\
& $\tau$-model
& $2.38-0.74$
& $1.64$
& $0.82$\\
\hline
\lw{3-jet}
& ${\rm CF_I(Exp)}$
& $3.09-0.63$
& $2.46$
& $1.23$\\
& $\tau$-model
& $3.02-0.77$
& $2.25$
& $1.13$\\
\hline
\end{tabular}
\end{table}

\paragraph{C5)}
As is seen in Table~\ref{tab5}, the Lorentz and Gaussian distributions are calculated from the Levy canonical form $P_4(\xi;1,0)$ and $P_4(\xi;2,0)$, respectively. Moreover, the asymptotic behavior in the $\xi$-space in our approach is calculated as follows:
\begin{eqnarray}
2\pi^2\xi^3\frac{1}{(1+(\xi/R)^2)^{5/2}}\ \mapright{\xi\gg 1}\ \xi^{-2}
\label{eq13}
\end{eqnarray}
However, in the $\tau$-model, by use of the linear regression method, we have confirmed in the following:
\begin{eqnarray}
S_{\rm \tau,BE}(\xi)\ \mapright{\xi\gg 1}\ \xi^{-2\alpha_{\tau}-1}
\label{eq14}
\end{eqnarray}
This behavior ($2\alpha_{\tau}=0.875$: 2-jet) is exactly expected in the Levy canonical form. 

\paragraph{C6)}
We observe that the 3-jet BEC exhibits almost chaotic properties, provided that the theory of QO is applied. This indicates that it is an ideal event for the study of BEC. The phases of proceeded pions are completely randomized~\cite{Biyajima:1979ak,Biyajima:1990ku,HanburyBrown:1956bqd}.

\paragraph{D1)}
L3 Collaboration did not report the single ratios, $C_2^{\rm data}(Q) = N^{(2+:2-;Q)}/N^{(+-;Q)}$ and $C_2^{\rm MC}(Q) = N_{\rm MC}^{(2+:2-;Q)}/N_{\rm MC}^{(+-;Q)}$. Thus, in the present study we cannot analyze them. In the future, as DELPHI Collaboration did, L3 Collaboration would publish data on $R_{\rm 2\,data}$, $R_{\rm 2\,det}$ and $(R_{\rm 2\,gen}/R_{\rm 2\,gen\mathchar`-noBE})$, we could examine the role of ${\rm LRC_{(Gauss)}}$ in BEC. See Appendix~\ref{secA}. 

Moreover, concerning the systematic errors and uncertainties, we present calculation based on OPAL BEC at Z$^0$-pole~\cite{Acton:1991xb}, because of no data on ${\rm SR^{MC}}$~\cite{DELPHI:1992axf,DELPHI:1991aa}, in Appendix~\ref{secC}. According to the calculations using OPAL BEC data~\cite{Acton:1991xb}, we are able to understand the magnitude of the systematic error of estimated values with ${\rm LRC_{(Gauss)}}$.\\

\noindent
{\it Acknowledgments.} One of the authors (M.B.) would like to thank the colleagues of the Center for General Education at Shinshu University.

\appendix
\section{\label{secA}Data on $N_{\rm MC}^{(2+:2-)}/N_{\rm MC}^{(+-)}$ at $Z^0$-pole by DELPHI Collaboration}
\begin{figure}[H]
  \centering
  \includegraphics[width=0.48\columnwidth]{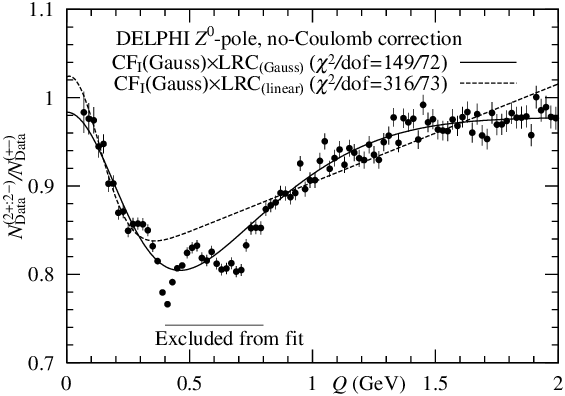}
  \includegraphics[width=0.48\columnwidth]{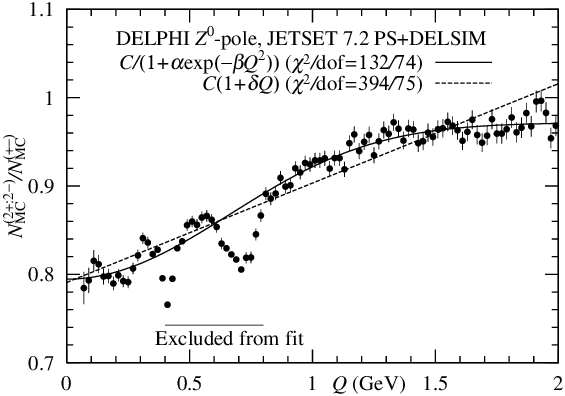}\\
  \includegraphics[width=0.48\columnwidth]{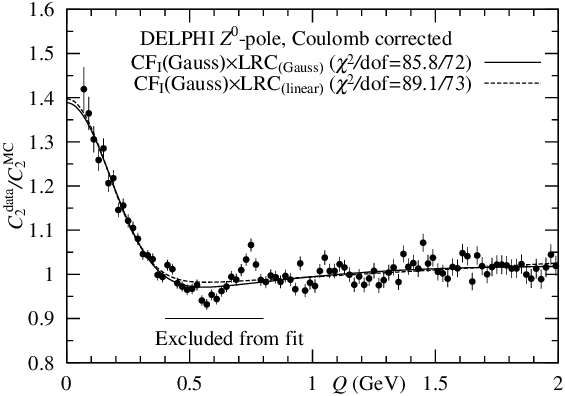}
  \includegraphics[width=0.48\columnwidth]{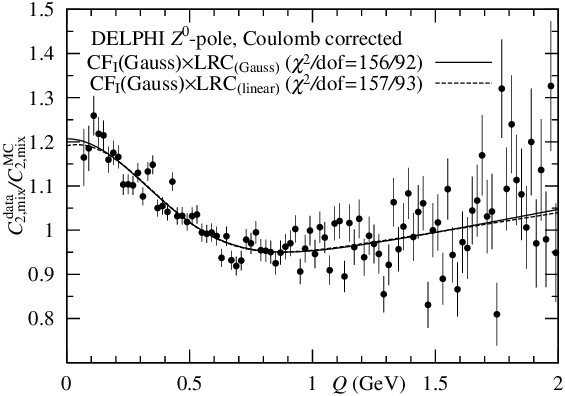}
  \caption{\label{fig6}Single ratios $N_{\rm data}^{(2+:2-)}/N_{\rm data}^{(+-)}$, $N_{\rm MC}^{(2+:2-)}/N_{\rm MC}^{(+-)}$, and the double ratios $C_2^{\rm data}/C_2^{\rm MC}$ and $C_{\rm 2,mix}^{\rm data}/C_{\rm 2,mix}^{\rm MC}$ shown by DELPHI Collaboration~\cite{DELPHI:1992axf} are analyzed by ${\rm CF_I(Gauss)\times LRC_{(Gauss)}}$ and ${\rm LRC_{(linear)}}$.}
\end{figure}

DELPHI Collaboration reported the data on $N_{\rm MC}^{(2+:2-)}/N_{\rm MC}^{(+-)}$ at $Z^0$-pole by making use of JETSET 7.2 with DELSIM~\cite{DELPHI:1992axf,Sjostrand:1986hx,DELPHI:1991aa}. The data are analyzed by ${\rm LRC_{(Gauss)}}$ and ${\rm LRC_{(linear)}}$ in Fig.~\ref{fig6}. ${\rm LRC_{(Gauss)}}$ seems to be better than ${\rm LRC_{(linear)}}$. About data in the interval $0.4<Q<0.8$ GeV, because of resonance effect from decays of $K^0$ and $\rho^0$ mesons, we do not use them in the present analysis. See also Ref.~\cite{DELPHI:1992axf}. 

The present results as shown in Figs.~\ref{fig6} imply that the ${\rm LRC_{(Gauss)}}$ is preferable to the ${\rm LRC_{(linear)}}$, except for $C_{\rm 2,mix}^{\rm data}/C_{\rm 2,mix}^{\rm MC}$. To confirm this statement, $\chi^2/$dof values (156/92 and 157/93) should be smaller.

In conclusion, we obtain the following results from analyses of ${\rm DR_{data}}=C_2^{\rm data}/C_2^{\rm MC}$ and ${\rm DR_{mix}}=C_{\rm 2,mix}^{\rm data}/C_{\rm 2,mix}^{\rm MC}$:
\begin{eqnarray*}
R_{\rm (Gauss)} &\!\!\!=&\!\!\! 0.62\pm 0.04\,({\rm stat})\pm 0.20\,({\rm sys})\ {\rm fm},\\
\lambda_{\rm (Gauss)} &\!\!\!=&\!\!\! 0.40\pm 0.03\,({\rm stat})\pm 0.05\,({\rm sys}).
\end{eqnarray*}

They are consistent with results shown in Ref.~\cite{DELPHI:1992axf}. When we assume the ${\rm CF_I(Exp)\times LRC_{(linear)}}$ for two DR's, the following results are obtained:
\begin{eqnarray*}
R_{\rm (Exp)} &\!\!\!=&\!\!\! 0.91\pm 0.05\,({\rm stat})\pm 0.48\,({\rm sys})\ {\rm fm},\\
\lambda_{\rm (Exp)}  &\!\!\!=&\!\!\! 0.93\pm 0.09\,({\rm stat})\pm 0.07\,({\rm sys}).
\end{eqnarray*}

\begin{table}[H]
\centering
\caption{\label{tab8}Fit parameters of data by DELPHI Collaboration by ${\rm CF_I(Gauss)\times LRC}$ and ${\rm CF_I(Exp)\times LRC_{(linear)}}$}
\vspace{1mm}
\begin{tabular}{cccccc}
\hline
${\rm CF_I(Gauss)}$
& LRC 
& $R$ (fm) 
& $\lambda$ 
& $\delta$ (GeV$^{-1}$)/($\alpha$, $\beta$ (GeV$^{-2}$)) 
& $\chi^2$/dof\\
\hline

$C_{2}^{\rm data}(Q)$
& linear
& $1.08 \pm 0.04$
& $0.29 \pm 0.02$
& $(14.1\pm 0.4)\times 10^{-2}$
&  316/73\\

no-Coulomb correction
& Gauss
& $0.69 \pm 0.04$
& $0.36 \pm 0.03$
& ($0.35 \pm 0.03$, $1.59 \pm 0.14$)
&  149/72\\
\hline

\lw{$C_{2}^{\rm MC}(Q)$}
& linear
& ---
& ---
& $(14.2\pm2  0.3)\times 10^{-2}$
&  394/75\\

& Gauss
& ---
& ---
& ($0.223\pm 0.004$, $1.42\pm  0.06$)
&  131/74\\
\hline

$C_{2}^{\rm data}(Q)/C_{2}^{\rm MC}(Q)$
& linear
& $0.83 \pm 0.03$
& $0.45 \pm 0.02$
& $(3.3\pm 0.7)\times 10^{-2}$
&  89.1/73\\

Coulomb correction
& Gauss
& $0.78 \pm 0.05$
& $0.47 \pm 0.03$
& ($0.075\pm 0.024$, $1.13 \pm 0.51$)
&  85.8/72\\
\hline

$C_{\rm 2,mix}^{\rm data}(Q)/C_{\rm 2,mix}^{\rm MC}(Q)$
& linear
& $0.42\pm 0.02$
& $0.38\pm 0.04$
& $0.10\pm 0.03$
&  157/93\\

Coulomb correction
& Gauss
& $0.42\pm 0.03$
& $0.33\pm 0.02$
& ($0.32\pm 0.03$, $0.20\pm 0.05$)
&  156/92\\
\hline
\hline

${\rm CF_I(Exp)}$
& LRC 
& $R$ (fm) 
& $\lambda$ 
& $\delta$ (GeV$^{-1}$)
& $\chi^2$/dof\\
\hline

$C_{2}^{\rm data}(Q)/C_{2}^{\rm MC}(Q)$
& linear
& $1.38 \pm 0.08$
& $0.86 \pm 0.05$
& $0.04\pm 0.01$
&  107/73\\

$C_{\rm 2,mix}^{\rm data}(Q)/C_{\rm 2,mix}^{\rm MC}(Q)$
& linear
& $0.43\pm 0.02$
& $1.00\pm 0.12$
& $0.28\pm 0.01$
&  161/93\\
\hline

\end{tabular}
\end{table}

\section{\label{secB}Levy canonical form and formulation obtained by the inverse Fourier transformation ($N=1$ and $N=4$)}
\begin{figure}[H]
  \centering
  \includegraphics[width=0.6\columnwidth]{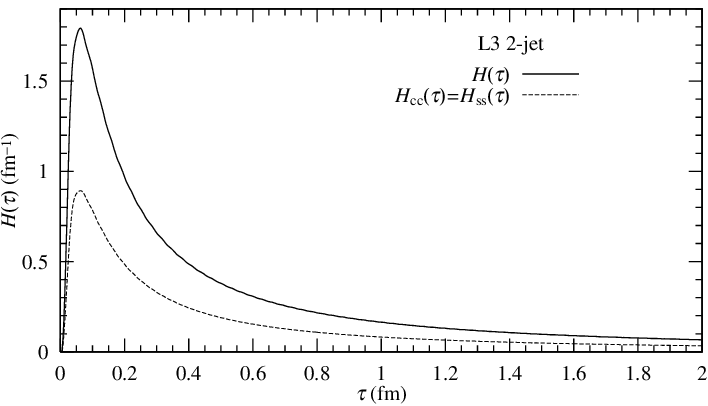}
  \caption{\label{fig7}The proper time function $H(\tau)$ of the Levy canonical form $\tilde H(\omega)$ introduced by L3 Collaboration.}
\end{figure}

\paragraph{1)}
First, we show the proper time function $H(\tau=\sqrt{t^2-r_z^2})$ calculated by L3 Collaboration~\cite{L3:2011kzb}. According to the notation in Ref.~\cite{L3:2011kzb}, the Levy canonical form with $\zeta = \tan(\alpha_{\tau}\pi/2)$ is expressed as
\begin{eqnarray}
\tilde H(\omega) = \exp\left[-\frac 12(\Delta r\omega)^{\alpha_{\tau}}(1-i\,{\rm sign}(\omega)\zeta)\right].
\label{eq15}
\end{eqnarray}
The function $H(\tau)$ is calculated by the inverse Fourier transformation as 
\begin{eqnarray}
H(\tau) &=& \frac 1{\pi}\int_0^{\infty}e^{-i\omega\tau}\tilde H(\omega)d\omega
\nonumber\\ 
 &=& \frac 1{\pi}{\rm Re}\left[\int_0^{\infty} \exp\left(-\frac 12(\Delta r\omega)^{\alpha_{\tau}}-i\,\omega\tau + \frac i2(\Delta r\omega)^{\alpha_{\tau}}\zeta\right)d\omega\right]
\nonumber\\
 &=& \frac 1{\pi}\int_0^{\infty} \exp\left(-\frac 12(\Delta r\omega)^{\alpha_{\tau}}\right)\cos\left(\omega\tau - \frac 12(\Delta r\omega)^{\alpha_{\tau}}\zeta\right)d\omega
\nonumber\\
 &=& H_{\rm cc}(\tau)+H_{\rm ss}(\tau)
\label{eq16}
\end{eqnarray}
where cc and ss represent the products of $\cos\cdot\cos$ and $\sin\cdot\sin$ respectively. Using $\Delta r = 1.56$ fm and $\alpha_{\tau}=0.44$, $H(\tau)$, $H_{\rm cc}(\tau)$ and $H_{\rm ss}(\tau)$ are calculated (Fig.~\ref{fig7}). The magnitudes of area presented in Fig.~\ref{fig7} are $S\approx 0.9$ and $S_{\rm cc}=S_{\rm ss}\approx 0.45$ in the range of $0\sim 50$ fm.

Combining the transverse and the rapidity distributions $\rho(p_T)$ and $\rho(y)$, L3 Collaboration calculated their emitting source functions.
%

\begin{figure}[H]
  \centering
  \includegraphics[width=0.6\columnwidth]{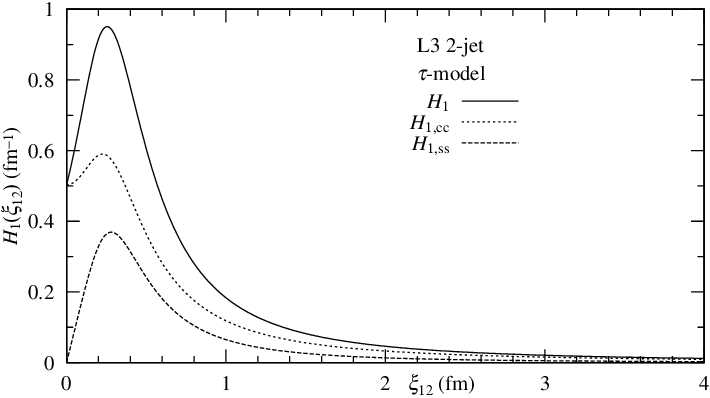}
  \caption{\label{fig8}The probability $H_{1}(\xi)$ of $\tilde H_{1}(\omega)$.}
\end{figure}

\paragraph{2)}
Second, to analyze the 2-jet events, they introduced the transverse mass $a=1/m_t$. With the following identification, $\Delta ra/2=R^2$, they started with the following Levy canonical form ($N=1$): 
\begin{eqnarray}
\tilde H_{1}(\omega) = \exp\left[-\frac 12(R\omega)^{2\alpha_{\tau}}(1-i\,{\rm sign}(\omega)\zeta)\right].
\label{eq17}
\end{eqnarray}
By making use the inverse Fourier transformation, the probability ($N=1$) $H_{1}(\xi_{12}=|x_1-x_2|)$ is calculated as
\begin{eqnarray}
H_{1}(\xi_{12}=|x_1-x_2|) = \frac 1{\pi}\int_0^{\infty} \exp\left[-\frac 12(R\omega)^{2\alpha_{\tau}}\right]\cos\left[\xi \omega - \frac 12(R\omega)^{2\alpha_{\tau}}\zeta\right]d\omega,
\label{eq18}
\end{eqnarray}
The results ($R=0.78$ fm, $\alpha_{\tau}=0.44$) are shown in Fig.~\ref{fig8}. Two figures in Fig.~\ref{fig7} and Fig.~\ref{fig8} are different each other, because of difference between $(\Delta r\omega)^{\alpha_{\tau}}$ and $(R\omega)^{2\alpha_{\tau}}$.
%
The magnitude of $H(\xi_{12})$, $S_1=\dint H(\xi_{12})d\xi_{12}\cong 0.75$ is computed in the range  $0\sim 40$ fm. Because its magnitude, the name of probability for $H(\xi_{12})$ seems to be unsuitable. Note that present calculation is based on $N=1$.

\paragraph{3)}
Third, L3 Collaboration adopted the following effective Levy form ($N=4$),
\begin{eqnarray}
\tilde H_{4}(R\omega\to RQ) = \exp\left[-\frac 12(RQ)^{2\alpha_{\tau}}(1-i\,{\rm sign}(Q)\zeta)\right].
\label{eq19}
\end{eqnarray}
From Eq.~(\ref{eq19}), we calculate the probability $P_{\tau}(\xi)$ as
\begin{eqnarray}
\left\{
\begin{array}{l}
H_4(\xi) = \dfrac 1{(2\pi)^2\xi} \dint Q_{\xi}^2\, \exp\left(-\frac 12(RQ_{\xi})^{2\alpha_{\tau}}\right)\cos\left(\frac 12(RQ_{\xi})^{2\alpha_{\tau}}\zeta\right)\, J_1(Q_{\xi}\xi)dQ_{\xi},
\medskip\\
P_{\tau}(\xi)=2\pi^2\xi^3 H_4(\xi),
\end{array}
\right.
\label{eq20}
\end{eqnarray}
where $\displaystyle{\int_0^{\pi}} \sin(\xi Q\cos\phi)\sin^2\phi\, d\phi=0$ is used.

\begin{figure}[H]
  \centering
  \includegraphics[width=0.6\columnwidth]{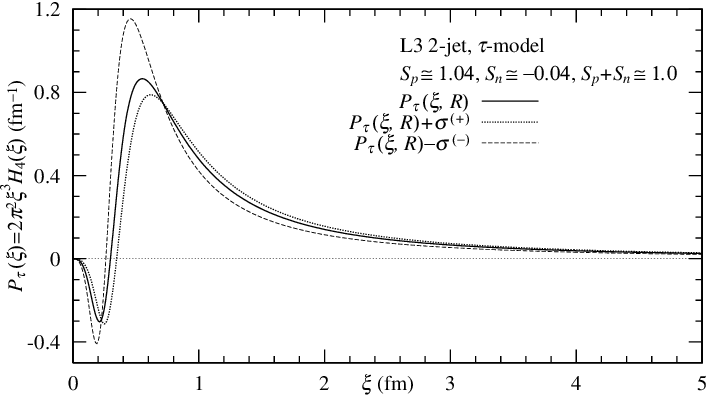}
  \caption{\label{fig9}The probability $P_{\tau}(\xi)$ calculated by Eq.~(\ref{eq20}). $\sigma^{(\pm)} = \partial P_{\tau}(\xi,\,R)/\partial R\cdot \delta R^{(\pm)}$ with $\delta R^{(+)}=\sqrt{(0.04(\rm stat))^2+(0.09(\rm sys))^2}$ and $\delta R^{(-)}=\sqrt{(0.04(\rm stat))^2+(0.16(\rm sys))^2}$. The values ($0.01_{-0.160}^{+0.09}$) are taken from Table 3 in Ref.~\cite{L3:2011kzb}.}
\end{figure}

The probability $P_{\tau}(\xi)$ ($R=0.78$ fm, $\alpha_{\tau}=0.44$) is shown in Fig.~\ref{fig9}. That the magnitude of $S=\int P_{\tau}(\xi)d\xi\cong 1.0$ is computed. However, it is very strange that $P_{\tau}(\xi)$ shows the partially negative behavior in the range of $0\sim 0.4$ fm: $S_p\cong 1.04$ and $S_n\cong -0.04$, which mean the positive and the negative magnitudes, respectively~\cite{Nolan:2020aa,Zolotarev:1986aa}.
%
The exchange function $E_{\rm BE}$ in the $\tau$-model is calculated as follows,
\begin{eqnarray}
E_{\rm BE} &=& {\rm Re}\left[\tilde H_{4}^2(RQ)\right]
\nonumber\\
 &=& \exp\left(-(RQ)^{2\alpha_{\tau}}\right)\cos\left((RQ)^{2\alpha_{\tau}}\zeta\right).
\label{eq21}
\end{eqnarray}
Introducing the degree of coherence $\lambda$ and ${\rm LRC_{(linear)}}$, we obtain Eq.~(\ref{eq2}). 

The stochastic density of exchange function $E_{\rm BE}$ in the $\tau$-model is calculated by the inverse Fourier transformation ($N=4$) as 
\begin{eqnarray}
\left\{
\begin{array}{l}
\mbox{\Large $\rho$}_{\rm \tau, BE}(\xi) = \dfrac 1{(2\pi)^2\xi} \dint Q_{\xi}^2\, \exp(-(RQ_{\xi})^{2\alpha_{\tau}})\cos((RQ_{\xi})^{2\alpha_{\tau}}\zeta)\, J_1(Q_{\xi}\xi)dQ_{\xi},
\medskip\\
S_{\rm \tau, BE}(\xi) = 2\pi^2\xi^3\mbox{\Large $\rho$}_{\rm \tau, BE}(\xi).
\end{array}
\right.
\label{eq22}
\end{eqnarray}
%

\section{\label{secC}Systematic errors and uncertainties}
\subsection*{I) BEC at Z$^0$-pole by OPAL Collaboration}
To explain the systematic errors, we need the various SRs and DRs in BEC by L3 Collaboration. However, because of no-information on various distributions by L3 Collaboration, we treat the SR and DR by OPAL Collaboration. Using estimated values in Table~\ref{tab9}, and the definition given in Ref.~\cite{Acton:1991xb}, we are able to calculate the systematic errors as 
\begin{eqnarray*}
&\!\!\!&\!\!\! \delta \lambda_{\rm sys}= \pm\sqrt{(\lambda_a-\lambda_b)^2+(\lambda_a-\lambda_c)^2+(\lambda_a-\lambda_d)^2}=\pm 0.120,\\ 
{\rm and}&\!\!\!&\!\!\! \delta R_{\rm sys}= \pm\sqrt{(R_a-R_b)^2+(R_a-R_c)^2+(R_a-R_d)^2}=\pm 0.260\ {\rm fm}.
\end{eqnarray*}
Finally, we obtain the following values:
\begin{eqnarray*}
\lambda = 0.736\pm 0.035\,({\rm stat})\pm 0.120\,({\rm sys}),\ {\rm and}\ 
R = 0.984\pm 0.029\,({\rm stat})\pm 0.260\,({\rm sys})\ {\rm fm}.
\end{eqnarray*}
Those values are compared with $\lambda = 0.866\pm 0.032\,({\rm stat})\pm 0.140\,({\rm sys})$, and $R = 0.928\pm 0.019\,({\rm stat})\pm 0.150\,({\rm sys})$ fm as ${\rm LRC_{(OPAL)}}$ is used~\cite{Acton:1991xb}. The absolute value $\delta R_{\rm sys}=0.260$ fm is larger than 0.150 fm~\cite{Acton:1991xb}.
%
\begin{table}[H]
\centering
\caption{\label{tab9}Fitting parameters in application of ${\rm CF_I(Gauss)\times LRC_{(Gauss)}}$ to data at $Z^0$-pole\\ by OPAL Collaboration. Notice that we have no-distribution with conditions ``different data selection'' and ``use 0.05 GeV binning''.}
\vspace{1mm}
\begin{tabular}{lccc}
\hline
data
& $\lambda$
& $R$ (fm)
& $\chi^2$/ndf\\
\hline
(a) SR excluding $K^0$ and $\rho^0$ decay effect
& $0.736\pm 0.035$
& $0.984\pm 0.029$
& 97.6/59\\

(b) SR full data
& $0.655\pm 0.025$
& $0.794\pm 0.029$
& 287/73\\

(c) SR $0<Q<1.5$ GeV
& $0.655\pm 0.026$
& $0.817\pm 0.033$
& 273/53\\

(d) DR full data
& $0.701\pm 0.038$
& $0.917\pm 0.032$
& 98.1/73\\
\hline
\end{tabular}
\end{table}

\begin{figure}[H]
  \centering
  \includegraphics[width=0.48\columnwidth]{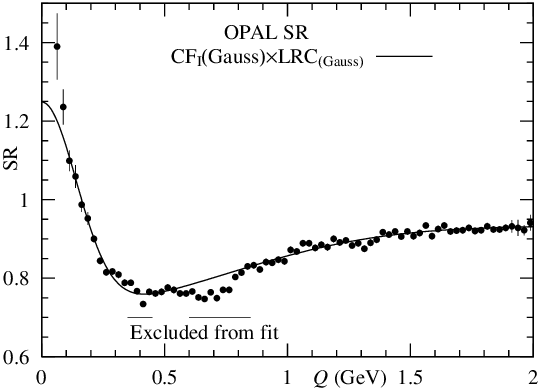}
  \includegraphics[width=0.48\columnwidth]{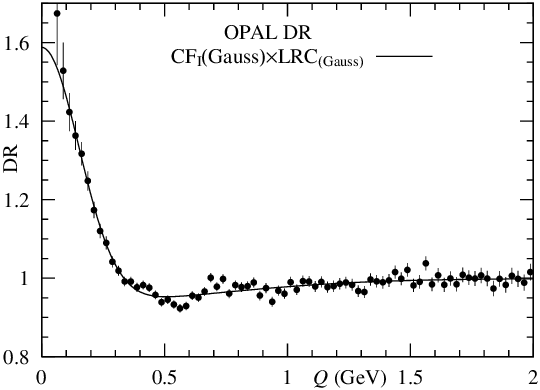}
  \caption{\label{fig10}Analysis of OPAL BEC data by Eq.~(\ref{eq6}).}
\end{figure}

When we use the different definition with the largest positive and negative differences among $\lambda$'s and $\lambda_{\rm (a)}$, and among $R$'s and $R_{\rm (a)}$, where the suffix (a) means the first line (a) in Table~\ref{tab9}, we obtain the following values $\delta \lambda_{\rm sys}=0.035_{-0.081}^{+0.0}$ and $\delta R_{\rm sys}=0.029_{-0.190}^{+0.0}$ fm for the systematic uncertainties.

\begin{figure}[H]
  \centering
  \includegraphics[width=0.48\columnwidth]{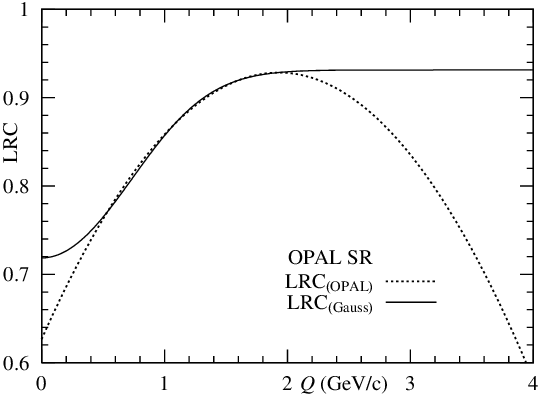}  
  \includegraphics[width=0.48\columnwidth]{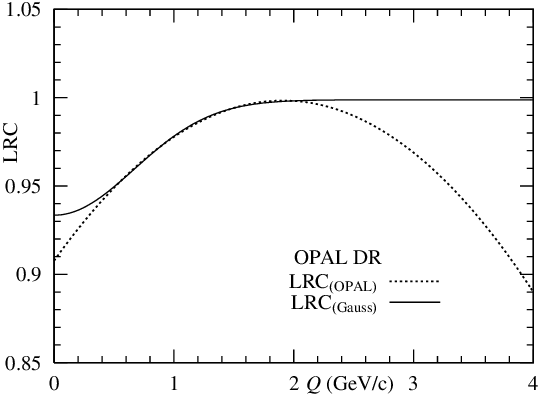}
  \caption{\label{fig11}${\rm LRC_{(OPAL)}}$ and ${\rm LRC_{(Gauss)}}$ for SR and DR by OPAL Collaboration.}
\end{figure}

In Fig.~\ref{fig11}, we show two behaviors related to ${\rm LRC_{(OPAL)}}$ and ${\rm LRC_{(Gauss)}}$ in SR and DR. The difference is observed in the range of $0<Q<0.5$ GeV. On the contrary, in the range of $0.5<Q<2.0$ GeV, the coincidences are seen.

\subsection*{II) BEC at Z$^0$-pole by L3 Collaboration}
Since we have no-information on the numerator and the denominator in Eq.~(\ref{eq2}), we use the systematic uncertainties in BEC at Z$^0$-pole by L3 Collaboration in Table 2 in Ref.~\cite{L3:2011kzb}:
$$
R=0.78\pm 0.04_{-0.16}^{+0.09}\ \mbox{fm for 2-jet}.
$$
These values are used in calculation of $\sigma^{\pm}$ in Fig.~\ref{fig5} and Fig.~\ref{fig9}.
%


\begin{thebibliography}{99}

\bibitem{L3:2011kzb} 
P.~Achard \textit{et al.} [L3],
Eur. Phys. J. C \textbf{71} (2011), 1648.

\bibitem{CMS:2019fur} 
A.~M.~Sirunyan \textit{et al.} [CMS],
JHEP \textbf{03} (2020), 014.

\bibitem{CMS:2011nlc} 
V.~Khachatryan \textit{et al.} [CMS],
JHEP \textbf{05} (2011), 029.

\bibitem{Acton:1991xb} 
  P.~D.~Acton {\it et al.} [OPAL],
  Phys.\ Lett.\ B {\bf 267} (1991) 143.

\bibitem{DELPHI:1992axf} 
P.~Abreu \textit{et al.} [DELPHI],
Phys. Lett. B \textbf{286} (1992), 201-210.

\bibitem{ALEPH:1991loh} 
D.~Decamp \textit{et al.} [ALEPH],
Z. Phys. C \textbf{54} (1992), 75-86.

\bibitem{Mizoguchi:2023zfu} 
T.~Mizoguchi, S.~Matsumoto and M.~Biyajima,
Phys. Rev. D \textbf{108} (2023) 5, 056014.

\bibitem{Mizoguchi:2021slz} 
T.~Mizoguchi, S.~Matsumoto and M.~Biyajima,
Int. J. Mod. Phys. A \textbf{37} (2022) 25, 2250148.

\bibitem{Shimoda:1992gb} 
R.~Shimoda, M.~Biyajima and N.~Suzuki,
Prog. Theor. Phys. \textbf{89} (1993), 697-708.

\bibitem{Wick:1954eu} 
G.~C.~Wick,
Phys. Rev. \textbf{96} (1954), 1124-1134.

\bibitem{Biyajima:1979ak} 
M.~Biyajima,
Phys. Lett. B \textbf{92} (1980), 193.

\bibitem{Biyajima:1990ku} 
M.~Biyajima, A.~Bartl, T.~Mizoguchi, O.~Terazawa and N.~Suzuki,
Prog. Theor. Phys. \textbf{84} (1990), 931-940.

\bibitem{Kozlov:2007dv} 
G.~A.~Kozlov, O.~V.~Utyuzh, G.~Wilk and Z.~Wlodarczyk,
Phys. Atom. Nucl. \textbf{71} (2008), 1502-1504.

\bibitem{HanburyBrown:1956bqd} 
R.~Hanbury Brown and R.~Q.~Twiss,
Nature \textbf{178} (1956), 1046-1048.

\bibitem{Goldhaber:1960sf} 
G.~Goldhaber, S.~Goldhaber, W.~Y.~Lee and A.~Pais,
Phys. Rev. \textbf{120} (1960), 300-312

\bibitem{Uchaikin:1999} 
V. V. Uchaikin, and  V. M. Zolotarev, 
``Chance and stability: Stable distributions and their applications,''
VSP, Utrecht, Netherlands, 1999:

\bibitem{Zolotarev:1981aa} 
V. M. Zolotarev, 
``Integral Transformation of Distributions and Estimates of Parameters of Multidimensional Spherically Symmetric Stable Laws'' in ``Contributions to Probability'' (Academic Press, New York 1981), pp. 283-305.

\bibitem{Zolotarev:1986aa} 
V. M. Zolotarev, 
``One-dimensional stable distributions,'' Translations of Mathematical Monographs Vol. 65,
American Mathematical Society, Providence, 1986.

\bibitem{Nolan:2020aa} 
J. P. Nolan, 
``Univariate Stable Distributions: Models for Heavy Tailed Data,''
Springer Nature, Switzerland, 2020.

\bibitem{Oberhettinger:1990} 
F. Oberhettinger,
``Tables of Fourier Transforms and Fourier Transforms of Distributions,''
Springer, Berlin/Heidelberg, 1990.

\bibitem{Gradshteyn:2014aa} 
I. S. Gradshteyn and I. ,M. Ryzhik, 
``Table of Integrals, Series, and Products,'' 8th edition,
Academic Press, Waltham, 2014.

\bibitem{Sjostrand:1986hx} 
T.~Sjostrand and M.~Bengtsson,
Comput. Phys. Commun. \textbf{43} (1987), 367.

\bibitem{DELPHI:1991aa} 
DELPHI Collaboration, 
``Comparison of data with QCD models,''
contributed paper to the LP-EPS conference, Geneva, August 1991.

\end{thebibliography}
\end{document}